# Transforming In-Vehicle Network Intrusion Detection: VAE-based Knowledge Distillation Meets Explainable AI

Muhammet Anıl Yağız
anill.yagiz@gmail.com
Kırıkkale University

Pedram MohajerAnsari
pmohaje@clemson.edu
Clemson University

Mert D. Pesé
mpese@clemson.edu
Clemson University

Polat Goktas
polat.goktas@ucd.ie
University College Dublin

## Abstract

In the evolving landscape of autonomous vehicles, ensuring robust in-vehicle network (IVN) security is paramount. This paper introduces an advanced intrusion detection system (IDS) called KD-XVAE that uses a Variational Autoencoder (VAE)-based knowledge distillation approach to enhance both performance and efficiency. Our model significantly reduces complexity, operating with just 1669 parameters and achieving an inference time of 0.3 ms per batch, making it highly suitable for resource-constrained automotive environments. Evaluations in the HCRL Car-Hacking dataset demonstrate exceptional capabilities, attaining perfect scores (Recall, Precision, F1 Score of 100%, and FNR of 0%) under multiple attack types, including DoS, Fuzzing, Gear Spoofing, and RPM Spoofing. Comparative analysis on the CICIoV2024 dataset further underscores its superiority over traditional machine learning models, achieving perfect detection metrics. We furthermore integrate Explainable AI (XAI) techniques to ensure transparency in the model's decisions. The VAE compresses the original feature space into a latent space, on which the distilled model is trained. SHAP (SHapley Additive exPlanations) values provide insights into the importance of each latent dimension, mapped back to original features for intuitive understanding. Our paper advances the field by integrating state-of-the-art techniques, addressing critical challenges in the deployment of efficient, trustworthy, and reliable IDSes for autonomous vehicles, ensuring enhanced protection against emerging cyber threats.

## 1 Introduction

The emergence of autonomous vehicles (AVs) and the widespread integration of connected cars have significantly transformed the landscape of modern transportation, offering unprecedented levels of safety, efficiency, and convenience. Central to these technological advancements is the in-vehicle network (IVN), which enables seamless communication among various electronic control units (ECUs) that govern essential vehicular functions, including adaptive cruise control, lane-keeping assistance, and real-time navigation [33, 34]. However, the increasing connectivity of IVNs introduces significant cyber vulnerabilities, primarily due to the lack of robust security measures such as authentication and encryption. These weaknesses can be exploited by malicious actors to carry out cyber attacks, compromising passenger safety and undermining the operational integrity of vehicles [17].

Intrusion Detection Systems (IDS) are essential for enhancing the security of IVNs. While traditional IDS approaches are somewhat effective, they often fail to address the sophisticated and evolving nature of cyber threats targeting IVNs. The dynamic nature of these attacks requires more advanced and flexible solutions. Recent advances in artificial intelligence (AI), particularly in machine learning (ML) and deep learning (DL), offer promising solutions to improve IVN security. These technologies can analyze vast amounts of network traffic data, learn intricate patterns of normal and anomalous behaviors, and detect previously unknown attack vectors in real time [16, 32].

However, deploying ML and DL-based IDS in IVNs presents several challenges. A significant issue is the black-box nature of these models, which obscures their decision-making processes and diminishes the trust of stakeholders in their reliability. This lack of transparency, along with the resource-intensive nature of these models, is particularly concerning in the automotive industry, where safety and sustainable computing are paramount [4, 7, 9, 24]. Existing studies reveal significant gaps in current IDS solutions for IVNs, particularly in providing clear explanations for detections, which is crucial for gaining stakeholder trust and ensuring regulatory compliance [3, 7]. Furthermore, the high computational demand and complexity of training these models pose practical challenges for real-time deployment in vehicles [17, 33]. These limitations underscore the need for innovative solutions that balance performance, explainability, and efficiency.

In this context, our paper introduces an advanced IDS for the Controller Area Network (CAN) — the most prominent IVN protocol — utilizing a variational autoencoder (VAE)-based knowledge distillation approach with an emphasis on sustainable and explainable AI. This method not only improves the efficiency and accuracy of anomaly detection but also ensures computational efficiency, making it suitable for deployment in resource-constrained environments. The primary contributions of this paper are as follows:

- **Improving Anomaly Detection in IVN Security with VAE-based Knowledge Distillation**: By employing VAEs and knowledge distillation techniques, we enhance the detection capabilities of the IDS, making it more effective in identifying subtle and complex anomalies within IVNs [33, 34].
- **Sustainable Computing (Time & Data Efficient)**: Our approach optimizes the computational resources required for training and deploying IDS models, making it feasible to implement in resource-constrained environments typical of autonomous vehicles [3, 17].
- **Enhancing Reliability of IVN Security Models**: By incorporating sustainable AI methods, we ensure that the IDS models are not only accurate but also efficient, thereby increasing the trust and reliability of the security system among stakeholders [7, 9].

- **Optimal Model Training Strategies for IVN Security using Knowledge Distillation in Deep Learning Models**: Our study explores optimal strategies for training deep learning models using knowledge distillation, ensuring that the IDS models are not only accurate but also interpretable and efficient [1, 4].
- **Explainability of IVN Security Models**: By integrating Explainable AI (XAI) methodologies, specifically SHapley Additive exPlanations (SHAP), we provide a comprehensive understanding of the model's decision-making processes. This transparency enhances stakeholder trust and allows for actionable insights into the security of IVN systems, reinforcing the robustness and reliability of our proposed framework [18].

This paper aims to advance IVN security by developing a robust, efficient, and sustainable IDS framework that addresses the unique challenges of modern autonomous vehicles. The integration of knowledge distillation into the IDS enhances its performance, providing a comprehensive solution for securing in-vehicle networks against emerging cyber threats.

Our proposed IDS framework, KD-XVAE, demonstrates remarkable performance across two distinct datasets. On the HCRL Car Hacking Dataset, our framework attains an F1 score, precision, recall, and accuracy of 1.0, accompanied by a low inference time of 0.3 milliseconds per batch. Similarly, on the CIC-IOV2024 dataset, KD-XVAE achieves perfect scores of 1.0 in F1 score, precision, recall, and accuracy. These outcomes represent a significant advancement over current methodologies. Furthermore, KD-XVAE surpasses traditional machine learning models on the CIC-IOV2024 dataset, underscoring its exceptional capability in accurately identifying and mitigating cyber threats within in-vehicle networks.

Notably, while there exists another proposed model that achieves perfect scores across all classes on the HCRL Car Hacking Dataset, KD-XVAE framework distinguishes itself through its superior parameter efficiency and reduced inference time per batch, making it a more practical and efficient solution for real-world applications.

The rest of the paper is structured as follows. Section 2 provides a comprehensive review of related work. Section 3 describes the proposed IDS framework, focusing on the integration of VAE-based knowledge distillation and sustainable AI practices. This section also covers the experimental setup, including the datasets and performance metrics used for the evaluation. Section 4 presents and compares the results of the ablation study of the proposed system with other state-of-the-art methods. Section 5 discusses shortcomings of this work and provides suggestions for future work before we concludes the paper in Section 6.

## 2 Related Work

This section provides a comprehensive review of the existing literature on security mechanisms designed to detect vulnerabilities and address security issues within autonomous vehicle IVNs. Additionally, it explores AI-based approaches for enhancing security within these networks.

### 2.1 Intrusion Detection Systems in IVNs

Intrusion Detection Systems (IDSes) are essential for securing networks such as the Controller Area Network (CAN) in vehicles, which inherently lack robust security features. Various strategies have been developed for detecting intrusions within CAN buses, employing different data analysis and ML techniques.

Rajapaksha *et al.* [27] categorized IDS approaches for in-vehicle networks into four primary types:

- Fingerprint-based (bus level)
- Parameter-monitoring-based (message level)
- Information-theoretic-based (data-flow level)
- Machine-learning-based (functional level)

Among these, the ML and DL methods have shown significant efficacy in identifying novel attacks on CAN buses through a comprehensive data analysis. Sun *et al.* [30] further classified intrusion detection algorithms into three groups: semantic-based, data-domain-based, and periodicity-based approaches. Semantic-based methods analyze the semantics of CAN messages. For instance, Zhao *et al.* [35] utilized an Artificial Neural Network (ANN) to enhance IDS precision. Data domain-based methods, such as those proposed by Qin *et al.* [26], predict CAN messages using long-short-term memory (LSTM) networks. Periodicity-based methods, exemplified by the LSTM-based approach from Hossain *et al.* [12], leverage the temporal patterns of CAN messages to detect anomalies, achieving high accuracy rates.

### 2.2 Integration of Advanced Techniques

Recent studies have integrated spatial and temporal features to enhance detection capabilities. Lo *et al.* [17] combined Convolutional Neural Networks (CNN) and LSTMs to successfully identify suspicious activities on the CAN bus. Despite their effectiveness, these models have limitations, such as susceptibility to replay attacks. Transfer learning has also been applied to IDS for IVNs. Yang *et al.* [34] demonstrated significant performance improvements using transfer learning in small data sets. Khatri *et al.* [14] showed that combining transfer learning with CNN-LSTM could significantly enhance performance through one-shot learning. To address the limitations of Recurrent Neural Networks (RNNs), such as vanishing and exploding gradient issues, attention mechanisms have been introduced. Similarly, Sun *et al.* [30] combined CNN-LSTM with attention to improve model robustness.

***Self-Attention Mechanisms and Transformer Architectures***. To overcome the challenges associated with RNNs, recent research has explored self-attention mechanisms. Nam *et al.* [21] utilized bidirectional Generative Pre-trained Transformer (GPT) for binary classification in vehicle IDS, while Nguyen *et al.* [23] proposed a transformer-based approach, outperforming traditional and DL models in both efficiency and precision. Gupta *et al.* [8] further improved transformer models by incorporating CNN in the embedding layer, improving prediction performance .

***Machine Learning for IVN Security***. Qayyum *et al.* [25] provided a comprehensive review of the challenges associated with the deployment of ML in vehicular networks, particularly focusing on security concerns in developing DL pipelines for connected and autonomous vehicles. Talpur and Gurusamy [31] evaluated the adoption of DL for automotive network security, analyzing various DL-based solutions and their effectiveness in intrusion detection, authentication, and privacy protection within vehicle communication systems. Gupta *et al.* [8] introduced an attention-enabled hierarchical deep neural network (AHDNN) to detect intrusions and ensure the safety of smart vehicles at both the node and the network levels. Ashraf *et al.* [2] utilized DL for intrusion detection in transportation systems, employing a LSTM autoencoder to identify malicious network behaviors in IVNs and vehicle-to-infrastructure networks. Mehedi *et al.* [20] proposed a LeNet-based deep transfer learning IDS model with active attribute selection to recognize and differentiate malicious CAN messages.

*Explainable AI for IVN Security.* While these methods have shown high effectiveness, their opaque black-box nature presents significant challenges. The lack of transparency and interpretability in their decision-making processes diminishes their reliability and trustworthiness, particularly from an industry standpoint. The automotive sector demands security solutions that not only achieve high detection accuracy, but also offer clear explanations for the detected anomalies. To meet this need, there is a growing focus on developing interpretable models. Madhav *et al.* [19] explored the use of explainable AI to improve decision making in autonomous vehicle systems. Houda *et al.* [13] proposed a framework incorporating explainable AI to improve the trustworthiness of deep learning techniques in intrusion detection. This framework helps cybersecurity experts systematically elucidate decision-making processes, thereby strengthening IoT network security. Renda *et al.* [28] examined the role of explainable AI in federated learning to advance technologies for next-generation 5G and 6G networks, with a focus on vehicle-to-everything applications.

Building on these cutting-edge methodologies, our study aims to develop an advanced IDS for vehicle CAN buses, using transformer architecture. This approach takes advantage of the latest advances in machine learning, deep learning, and sustainable AI to create a robust and efficient IDS capable of addressing the unique challenges posed by IVNs in autonomous vehicles.

## 3 Methodology

### 3.1 Threat Model

Understanding the specific characteristics and potential impacts of these attacks is essential to develop effective IDS. Each attack type exhibits unique patterns and behaviors that can be utilized to create robust detection algorithms. By incorporating comprehensive datasets such as the HCRL Car-Hacking [29] and CIC-IoV [22] datasets (see Section 4.1 for detailed description), researchers can simulate a diverse range of attack scenarios and assess the resilience of IDS against advanced cyber threats. In the following, we provide detailed descriptions of the attack types included in these popular datasets to elucidate their characteristics and potential impacts on vehicle security.

*Denial of Service (DoS).* A DoS attack aims to overwhelm the vehicle's CAN bus with an excessive number of messages. This flood of traffic can effectively block legitimate messages from being transmitted, disrupting normal vehicle operations. In a DoS scenario, the vehicle's ECUs may be unable to communicate critical information, leading to potential safety hazards and system failures. On the CAN bus, a DoS attack is very straightforward as CAN messages with low identifiers (IDs) have higher priority and can thus prevent messages with higher IDs from communicating on the bus. As a result, injecting a message with CAN ID 0x0 would block any legitimate traffic on the bus.

*Fuzzing.* Fuzzing attacks involve sending a large number of random, misformed, or unexpected messages to the CAN bus to discover vulnerabilities within the vehicle's ECUs. By exposing the system to a variety of unexpected inputs, fuzzing can reveal weaknesses that can be exploited to cause the ECUs to malfunction or crash, thereby compromising vehicle safety and functionality.

*Spoofing Attacks.* Spoofing attacks on vehicular systems involve the injection of falsified data into the CAN bus, disrupting the accuracy of critical system readings and undermining vehicle performance and safety. There is a myriad of vehicular signals on the CAN bus that is subject to spoofing attacks, with the most prominent being RPM, gear, gas, speed, and steering wheel angle [22, 29].

### 3.2 Data Pre-processing and Feature Scaling

Prior to training the IDS, rigorous data pre-processing steps were taken to ensure data quality and consistency. These steps included:

(1) **Missing Value Handling**: Any missing or incomplete data entries were identified and addressed by imputation techniques or by discarding incomplete records to maintain the integrity of the dataset.
(2) **Data Conversion**: The CAN IDs, originally in hexadecimal format, were converted to decimal to facilitate numerical proceWssing. This transformation was essential to standardize the data and ensure compatibility with machine learning algorithms.
(3) **Normalization**: To mitigate the effects of varying scales and units across different features, a Min-Max Scaler was employed. This scaling technique normalized the data within the range [0, 1], ensuring that all characteristics contributed equally to the learning process. This step was crucial for enhancing the convergence rate of the machine learning models and preventing any single feature from dominating the learning process. The Min-Max normalization formula is defined as follows:

$$x' = \frac{x - x_{\min}}{x_{\max} - x_{\min}} \tag{1}$$

where $x$ is the original value, $x'$ is the normalized value, $x_{\min}$ is the minimum value of the feature, and $x_{\max}$ is the maximum value of the feature.

### 3.3 Strategic Data Partitioning for Model Evaluation

To ensure a robust evaluation of the IDS, the datasets were divided into training and testing sets. In an unconventional approach aimed at rigorously testing the IDS, only 5% of the data was used for training, while the remaining 95% was reserved for testing. This split was chosen to simulate real-world scenarios where models must perform well even with limited training data. The split ensured that the model was tested on a diverse and extensive dataset, highlighting its generalization capabilities and robustness.

Table 1: Hyperparameter Configuration for CHD and CIC IOV Datasets

| Hyperparameter | Values for CHD | Values for CIC IOV |
| --- | --- | --- |
| Smooth factor | 1e-06 | 1e-08 |
| Number of encoder layers | 3 | 3 |
| Batch size | 32 | 32 |
| Learning rate | 0.0001 | 0.0001 |
| Epochs | 1 | 50 |

## 3.4 Proposed Model Architecture

To build a robust and efficient IDS, we employed a combination of advanced techniques, including VAEs and Knowledge Distillation.

*3.4.1 Variational Autoencoder (VAE).* The VAE architecture was chosen for its ability to learn latent representations of the data, capturing the underlying distribution of normal and anomalous patterns in IVNs. The VAE consisted of an encoder, which mapped the input data to a latent space, and a decoder, which reconstructed the input data from the latent space. The reconstruction error was then used to identify anomalies. The key parameters and structures of the VAE are summarized in Table 1.

***Encoder and Decoder Layers***. The encoder and decoder were designed with multiple dense layers, employing ReLU activations to capture complex non-linear relationships.

***Latent Space Dimension***. The dimension of the latent space was carefully selected to balance the trade-off between reconstruction fidelity and computational efficiency.

***Regularization***. KL-divergence regularization was applied to ensure the latent space followed a standard normal distribution, facilitating better generalization and anomaly detection.

*3.4.2 Knowledge Distillation.* Knowledge distillation, initially proposed by Hinton et al. [10], facilitates the transfer of knowledge from a large and complex model (referred to as *teacher*) to a smaller and simpler model (referred to as *student*). The core motivation behind knowledge distillation is to enable the student model to emulate the behavior of the teacher model, thereby achieving high accuracy with reduced computational demands.

***Proxy Dataset Selection***. The process of knowledge transfer typically requires a proxy dataset to act as the medium for transfer. Commonly, this proxy dataset is a mutually exclusive dataset; alternatively, some methods may employ an autoencoder or Generative Adversarial Network (GAN) to synthesize a suitable dataset.

***Soft Targets for Knowledge Transfer***. Once the proxy dataset is selected, the soft targets used for knowledge transfer are typically derived from the class probabilities of the teacher model's output layer (logits) or from feature representations within its intermediate layers (feature maps). These soft targets encapsulate more informative and nuanced representations compared to the hard labels utilized in conventional training paradigms.

***Loss Function***. In traditional knowledge distillation, the loss function comprises two components: a standard cross-entropy loss term and a distillation loss term. The distillation loss term is often defined as the Kullback-Leibler (KL) divergence between the softmax outputs of the teacher and student models:

$$L_{KL}(S \parallel T) = \sum_i S_i(\sigma(z)) \log\left(\frac{S_i(\sigma(z))}{T_i(\sigma(z))}\right) \quad (2)$$

where $S$ and $T$ represent the logits of the student and teacher models, $\sigma$ denotes the softmax function, and $T$ is a temperature parameter that smooths the probability distributions.

***Training Approach***. Our approach involves pre-training the teacher model on the full dataset to achieve high accuracy, generating soft labels from the teacher model's predictions, and then training the student model using a combination of hard and soft labels to inherit the knowledge of the teacher model while remaining computationally efficient.

Using knowledge distillation, our IDS achieves a balance between high detection accuracy and reduced computational resource requirements, making it suitable for real-time applications in resource-constrained environments.

*3.4.3 Explainable AI (XAI).* Interpretability is a fundamental aspect of our framework, enhancing trust and providing deep insight into the decision-making processes of our models. To illuminate the predictions made by the student model trained on the latent space representations from the VAE, we employ XAI techniques, focusing specifically on Shapley Additive Explanations (SHAP). SHAP values provide a consistent measure of feature importance, ensuring interpretability at both local and global levels [5].

The SHAP framework computes the Shapley values for each feature, quantifying their contributions to the model's predictions. The Shapley value for a feature is calculated as follows [18]:

$$\phi_i = \sum_{S \subseteq N \setminus \{i\}} \frac{|S|!(|N|-|S|-1)!}{|N|!} \left[v(S \cup \{i\}) - v(S)\right] \quad (3)$$

where $\phi_i$ represents the Shapley value for feature $i$, $N$ is the set of all features, $S$ denotes a subset of features and $v(S)$ is the value function reflecting the prediction of the model with subset $S$.

In our implementation, the VAE is utilized to compress the original high-dimensional feature space into a lower-dimensional latent space. The teacher/student model is then trained on these latent representations. To ensure transparency and comprehensibility of the model's decisions, we apply SHAP to the latent space. SHAP values indicate the importance of each latent dimension in the model's predictions. In addition, we map these latent dimensions back to the original features to provide a more intuitive understanding of the behavior of the model. This mapping, though simplified, offers valuable insights into how original features influence the latent space and, consequently, the model's predictions.

However, during the implementation of XAI techniques, we faced significant challenges related to memory consumption and processing time. The computational intensity of these methods necessitated the creation of a smaller, balanced subset of our dataset to enable more efficient analysis. This subset, consisting of 245,919 instances, was carefully curated to ensure an equal representation of each data category, thereby preserving the integrity and representatives of the original dataset. Despite the reduced size, KD-XVAE achieved an impressive macro average F1 score of 0.99, demonstrating the effectiveness of our approach and the robustness of the subset in maintaining essential data characteristics.

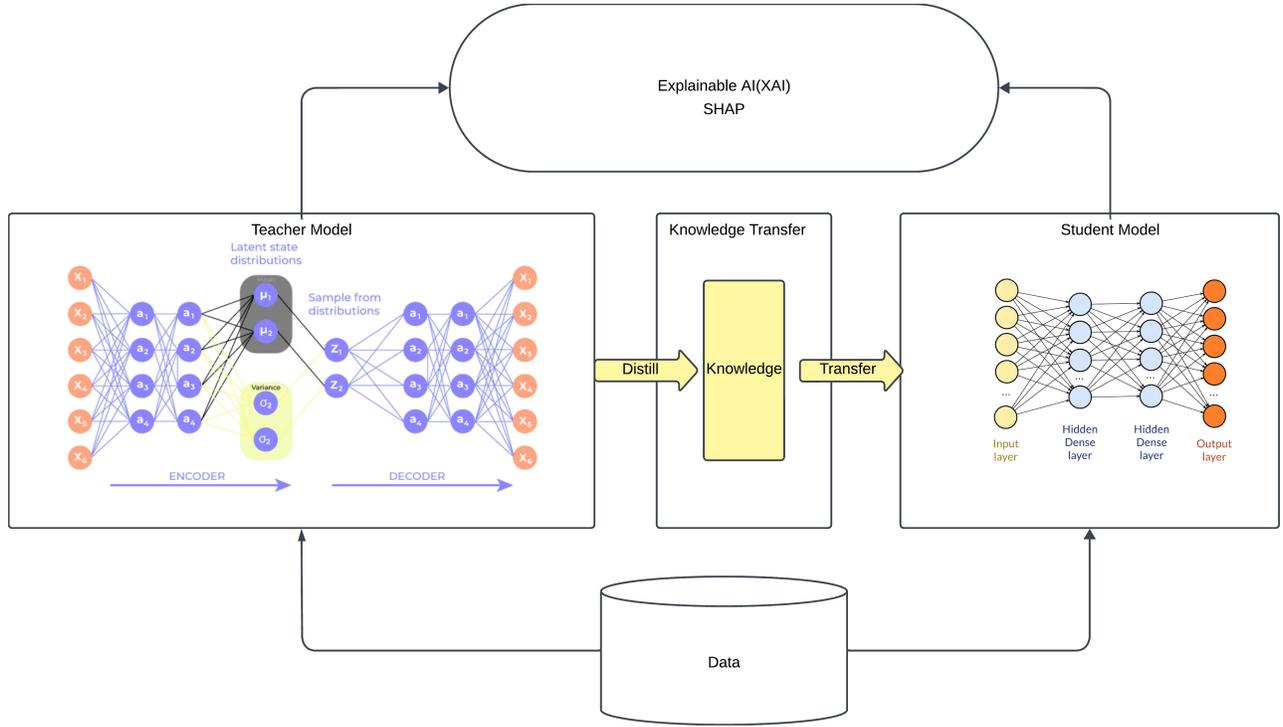

Figure 1: Proposed Model Architecture

By integrating SHAP into our framework, we not only improve the applicability of the model in real-world scenarios, but also provide stakeholders with actionable insights into the security and robustness of IoV systems. Thus, our framework enables both the reliability and interpretability necessary for effective deployment in practical applications.

## 4 Evaluation

In evaluating the performance of our proposed model against established competitors, we observed significant improvements in both efficiency and accuracy metrics.

### 4.1 Datasets

In this paper, two primary datasets were utilized to evaluate the performance of the IDS for IVNs, as detailed below:

***HCRL Car-Hacking***. Collected by the Hacking and Countermeasure Research Lab (HCRL), this dataset is a comprehensive resource for automotive cybersecurity. It includes both benign CAN traffic and four distinct types of *attacks—Denial of Service (DoS), Fuzzing, RPM Spoofing, and Gear Spoofing*. The data were gathered from a Hyundai YF Sonata vehicle under real-world conditions, providing a robust foundation for IDS development. Each CAN message includes a timestamp, an 11-bit CAN ID in HEX format (converted to DEC), data length, 8-byte data, and a flag indicating whether the message is normal or injected. Sequential samples ($n = 29$) are aggregated to form data frames, labeled according to the attack type if any injected message is present [29].

***Canadian Institute for Cybersecurity Internet of Vehicles (CIC-IoV)***. This dataset features DoS attacks and various spoofing scenarios conducted on a 2019 Ford vehicle. Attacks were executed under controlled conditions with the vehicle stationary, ensuring safety. It encompasses *DoS, Gas Spoofing, Steering Wheel Spoofing, Speed Spoofing, and RPM Spoofing*. Each entry contains a timestamp, CAN ID, Data Length Code (DLC), data bytes, and a flag indicating normal or injected messages. Sequential data frames are created by aggregating samples ($n = 7$), facilitating effective analysis and anomaly detection [22].

These datasets provide a comprehensive foundation for the development and evaluation of IDS in IVNs, enhancing the security and resilience of autonomous vehicles against sophisticated cyber threats, as summarized in Table 2.

### 4.2 Experimental Setup and Evaluation Metrics

The experiments are executed on a system featuring an Intel Core i9-13900K processor clocked at 3.0 GHz, 64 GB of DDR4 RAM, 1 TB SSD storage, and an NVIDIA GeForce RTX 3090 GPU with 24 GB of VRAM running Ubuntu 20.04 LTS.

In this study, we conducted comprehensive experiments to evaluate the performance of our proposed IDS model. The experiments were designed to rigorously test the model's ability to detect various types of cyber attack on IVNs. The evaluation was carried out using standard performance metrics and baseline comparisons with existing state-of-the-art IDS solutions. The key metrics used in this study are defined as follows:

- **Performance Metrics**: Metrics such as accuracy, precision, recall, and $F_1$-score were calculated to assess the model's detection capabilities. These metrics provided a holistic view of the model's effectiveness in identifying both normal and anomalous CAN messages.
- **Number of Parameters**:

$$\text{Number of Parameters} = \sum_{i=1}^{L} \text{Parameters}(i) \quad (4)$$

where $L$ is the number of layers in the model, and Parameters($i$) represents the number of parameters in the $i$-th layer.
- **Inference Time per Batch**:

$$\text{Inference Time (ms)} = \left( \frac{\text{End Time} - \text{Start Time}}{\text{Batch Size}} \right) \quad (5)$$

Table 2: Datasets Used for Vehicle Attack Analysis

| Dataset | Vehicle | Attacks Included | Conditions | Data Features | Number of Samples |
| --- | --- | --- | --- | --- | --- |
| HCRL Car-Hacking [29] | Hyundai YF Sonata | DoS<br>Fuzzing<br>RPM Spoofing<br>Gear Spoofing | Real conditions | Timestamp<br>CAN ID (HEX to DEC)<br>Data Length<br>Data bytes<br>Normal/Injected Flag<br>Sequential samples (n=29) | ~16,558,462 |
| CIC-IoV [22] | 2019 Ford vehicle | DoS<br>Gas Spoofing<br>Steering Wheel Spoofing<br>Speed Spoofing<br>RPM Spoofing | Controlled conditions | Timestamp<br>CAN ID (HEX to DEC)<br>DLC<br>Data bytes<br>Normal/Injected Flag<br>Sequential samples (n = 7) | ~1,408,219 |

Inference time per batch measures the time taken by the model to make predictions on a batch of input data. This metric indicates the efficiency and speed of the model during inference.

These comprehensive baseline comparisons illustrate the robustness, efficiency, and advanced capabilities of our proposed IDS model, establishing it as a leading solution for enhancing IVN security in autonomous vehicles.

### 4.3 Performance Evaluation on HCRL Car-Hacking Dataset

Table 4 provides a comprehensive comparison of performance metrics for various IDS models in the HCRL Car Hacking dataset, covering different attack types such as *DoS, Fuzzing, Gear Spoofing, and RPM Spoofing*. The detailed results are outlined below:

- **DoS attacks**: KD-XVAE achieves an FNR of 0%, matching the perfect performance of the Transformer [23] and Time-Embedded Transformer [15]. The Recall, Precision, and F1 Score for KD-XVAE are all 1.0, indicating flawless detection and classification of DoS attacks. This performance is on par with the best models but with a significantly lower computational footprint.
- **Fuzzing attacks**: KD-XVAE again demonstrates a perfect performance with an FNR of 0%, and Recall, Precision, and F1 Scores of 1.0. This surpasses the performance of models like Rec-CNN [6], which shows a high FNR of 17.0% and lower recall and precision values. The superiority of KD-XVAE in handling Fuzzing attacks highlights its robustness in identifying subtle and complex anomalies.
- **Gear Spoofing attacks**: KD-XVAE maintains an FNR of 0% and perfect scores in Recall, Precision, and F1 Score. This is a significant achievement compared to models like the LSTM [23] and CNN-LSTM [23], which have lower accuracy metrics. The high performance across all metrics confirms KD-XVAE's ability to effectively detect and classify Gear Spoofing attacks.
- **RPM Spoofing attacks**: KD-XVAE consistently delivers an FNR of 0%, with Recall, Precision, and F1 Score all at 1.0. This performance is comparable to the Time-Embedded Transformer [15] and superior to traditional models like DCNN ResNet [29], which, despite high precision, have lower recall values and higher FNRs.

In essence, KD-XVAE achieves perfect detection metrics (Recall, Precision, F1 Score of 1.0 and FNR of 0%) across all attack types in the HCRL Car-Hacking dataset. Performance consistency is particularly evident when compared to other models such as the Transformer [23] and the Time-Embedded Transformer [15], both of which also achieved high metrics but with greater computational demands.

### 4.4 Comparative Analysis on CICIoV2024 Dataset

Table 5 highlights the performance of KD-XVAE compared to traditional machine learning models on the CICIoV2024 dataset. The results clearly demonstrate the superiority of KD-XVAE in terms of accuracy and detection capabilities.

- **Accuracy**: KD-XVAE achieves a perfect accuracy score of 1.0, surpassing Logistic Regression (0.89), AdaBoost (0.92), Deep Neural Network (0.96), and Random Forest (1.0) [22]. This indicates KD-XVAE's ability to classify all instances correctly.

Table 3: Comparison of Model Complexity and Inference Time per Batch

| Model | Parameters (millions) | Inference time / batch (ms) |
| --- | --- | --- |
| Time-embedded Transformer with AutoEncoder [15] | 0.259 | 3.58 |
| Transformer [23] | 2.67 | 11.6 |
| Time-embedded Transformer [15] | 0.256 | 3.42 |
| Teacher Model | 0.00435 | 0.3 |
| Student Model | **0.0016** | **0.3** |

- **Recall**: With a recall of 1.0, KD-XVAE detects all attack instances. In comparison, Logistic Regression and AdaBoost show recalls of 0.50 and 0.66, respectively, while Deep Neural Network and Random Forest achieve 0.76 and 1.0.
- **Precision**: KD-XVAE's precision is also 1.0, indicating no false positives. Traditional models like Logistic Regression and AdaBoost have lower precision (0.48 each), whereas Deep Neural Network and Random Forest show 0.83 and 1.0, respectively.
- **F1 Score and Balanced Performance**: KD-XVAE's F1 scores of 1.0, reflecting the harmonic mean of precision and recall, were consistently perfect, showcasing a balanced performance in detecting intrusions while maintaining high precision and recall rates. Logistic Regression and AdaBoost have lower F1 scores of 0.49 and 0.51, respectively, while Deep Neural Network and Random Forest score 0.78 and 1.0. This balance is essential for effective and reliable IDS performance in dynamic and varied IVN environments.

Thus, KD-XVAE outperforms traditional machine learning models in the CICIoV2024 data set across all key performance metrics, showcasing its exceptional capability in detecting and classifying intrusions accurately and efficiently. These results validate the efficacy and superiority of our advanced deep learning approach in enhancing the security of in-vehicle networks.

## 4.5 Model Complexity and Inference Time

KD-XVAE's architecture, designed to be more compact, utilizes only 1669 parameters provided in Table 3. This is a considerable reduction compared to the Time-Embedded Transformer with Autoencoder, which operates with 259k parameters [15]. The streamlined parameter count directly correlates with reduced computational overhead, making KD-XVAE more suitable for resource-constrained environments typical of autonomous vehicles.

In terms of inference time, KD-XVAE demonstrates exceptional performance, requiring only 0.3 ms per batch, measured on the HCRL Car-Hacking dataset. This is a substantial improvement over the 3.58 ms needed by the Time-Embedded Transformer with Autoencoder [15]. Such efficiency gains are critical for real-time applications, ensuring prompt detection and response to cyber threats. Minimum cycle times for CAN messages on a 500 kBit/s bus as used for the two datasets stand at 10 ms [24]. As a result, any latency overhead caused by KD-XVAE needs to be smaller than 10 ms which also includes network latency and regular computation time. Our numbers easily satisfy this requirement, displaying the real-time capability of KD-XVAE.

## 4.6 XAI Analysis

To further understand the inner workings of KD-XVAE and ensure transparency, we conducted an XAI analysis to identify the most and least influential features and latent variables related to different classes and datasets. The insights gained from this analysis are crucial for validating model decisions and ensuring trustworthiness in critical applications like automotive cybersecurity.

*4.6.1 Latent Variables and Feature Influence in HCRL Car-Hacking Dataset.* In the HCRL Car-Hacking dataset, KD-XVAE's latent dimensions are influenced by specific features as follows:

- **Latent dimension 0, 11, 22**: Influenced by Timestamp.
- **Latent dimension 1, 12, 23**: Influenced by CAN ID.
- **Latent dimension 2, 13, 24**: Influenced by DLC.
- **Latent dimension 3-10, 14-21, 25-31**: Influenced by DATA[0] to DATA[7].

This detailed mapping indicates that the model relies heavily on data-related features to detect various types of attack. For example, the perfect detection of DoS attacks can be attributed to the influential role of CAN ID and DATA fields, which are critical in identifying anomalies in data transmission.

*4.6.2 Latent Variables and Feature Influence in CIC-IoV Dataset.* Similarly, in the CIC-IoV dataset, the influential features for the latent dimensions are as follows:

- **Latent dimension 0, 9, 18, 27**: Influenced by CAN ID.
- **Latent dimension 1-8, 10-17, 19-26, 28-31**: Influenced by DATA[0] to DATA[7].

Here, the importance of CAN ID and DATA fields is evident across multiple latent dimensions, reflecting the model's focus on these features for accurate classification.

*4.6.3 Feature Importance and Class-Specific Analysis.* The XAI analysis provides a detailed understanding of the importance of features in different classes in the HCRL Car Hacking and the CIC-IoV dataset.

- **HCRL Car-Hacking Dataset**:
  - **High Importance**:
    * CAN ID: Crucial for detecting DoS, Fuzzing, Gear Spoofing, and RPM Spoofing attacks.
    * DATA Fields: All data bytes (DATA[0] to DATA[7]) are highly influential in identifying DoS, Fuzzing, Gear Spoofing, and RPM Spoofing attacks.
  - **Low Importance**:
    * Timestamp: Low significance in identifying attack-free instances (only for Attack-Free class).

Table 4: Performance Metrics Comparison for Different IDS Models on the HCRL Car-Hacking Dataset Across Various Attack Types

| Attack Type | Model | FNR (%) | Recall | Precision | F1 Score |
|---|---|---|---|---|---|
| DoS | DCNN ResNet [29] | 0.3 | 0.9970 | 0.9999 | 0.9985 |
| | CE ResNet [11] | 0.21 | 0.9979 | 0.9997 | 0.9987 |
| | Rec-CNN [6] | 2.0 | 0.9800 | 0.9924 | 0.9862 |
| | LSTM [23] | 0.74 | 0.9926 | 0.9911 | 0.9919 |
| | CNN-LSTM [23] | 0.5 | 0.9945 | 0.9994 | 0.9969 |
| | Transformer [23] | **0** | **1.0** | **1.0** | **1.0** |
| | Time-Embedded Transformer [15] | **0** | **1.0** | **1.0** | **1.0** |
| | **KD-XVAE** | **0** | **1.0** | **1.0** | **1.0** |
| Fuzzing | DCNN ResNet [29] | 0.33 | 0.9967 | 0.9992 | 0.9979 |
| | CE ResNet [11] | 0.24 | 0.9976 | 0.9997 | 0.9987 |
| | Rec-CNN [6] | 17.0 | 0.8300 | 0.8964 | 0.8619 |
| | LSTM [23] | 3.27 | 0.9673 | 0.9666 | 0.9670 |
| | CNN-LSTM [23] | 0.9 | 0.9908 | 0.9991 | 0.9949 |
| | Transformer [23] | 0.01 | 0.9999 | 0.9999 | 0.9999 |
| | Time-Embedded Transformer [15] | **0** | **1.0** | **1.0** | **1.0** |
| | **KD-XVAE** | **0** | **1.0** | **1.0** | **1.0** |
| Gear Spoofing | DCNN ResNet [29] | 0.2 | 0.9980 | 0.9996 | 0.9988 |
| | CE ResNet [11] | 0.11 | 0.9992 | 0.9994 | 0.9997 |
| | Rec-CNN [6] | 7.63 | 0.9237 | 0.824 | 0.8763 |
| | LSTM [23] | 3.12 | 0.9688 | 0.9683 | 0.9685 |
| | CNN-LSTM [23] | 0.78 | 0.9922 | 0.9968 | 0.9945 |
| | Transformer [23] | 0.32 | 0.9968 | 0.9938 | 0.9953 |
| | Time-Embedded Transformer [15] | **0** | **1.0** | **1.0** | **1.0** |
| | **KD-XVAE** | **0** | **1.0** | **1.0** | **1.0** |
| RPM Spoofing | DCNN ResNet [29] | 0.19 | 0.9982 | 0.9995 | 0.9988 |
| | CE ResNet [11] | 0.12 | 0.9994 | 0.9996 | 0.9995 |
| | Rec-CNN [6] | 15.50 | 0.9345 | 0.8547 | 0.8923 |
| | LSTM [23] | 2.50 | 0.9726 | 0.9701 | 0.9713 |
| | CNN-LSTM [23] | 0.6 | 0.9912 | 0.9907 | 0.9909 |
| | Transformer [23] | 0.06 | 0.9998 | 0.9998 | 0.9998 |
| | Time-Embedded Transformer [15] | **0** | **1.0** | **1.0** | **1.0** |
| | **KD-XVAE** | **0** | **1.0** | **1.0** | **1.0** |

* DLC (Data Length Code): Low importance in distinguishing normal traffic (only for Attack-Free class).
- **CIC-IoV Dataset**:
  – **High Importance**:

* CAN ID: Identifies the IDs involved in DoS, RPM spoofing, Speed spoofing, Steering wheel spoofing, and Gas spoofing.

Table 5: Performance Metrics Comparison for Different ML Models on the CIC-IoV Dataset.

| Metrics | Logistic Regression [22] | AdaBoost [22] | Deep Neural Network [22] | Random Forest [22] | KD-XVAE |
|---|---|---|---|---|---|
| Accuracy | 0.89 | 0.92 | 0.95 | 0.96 | **1.0** |
| Recall | 0.50 | 0.66 | 0.76 | 0.76 | **1.0** |
| Precision | 0.48 | 0.48 | 0.83 | 0.76 | **1.0** |
| F1-score | 0.49 | 0.51 | 0.78 | 0.76 | **1.0** |

- * DATA Fields: Reflects payload patterns indicative of DoS Attack, RPM Spoofing, Speed Spoofing, Steering Wheel Spoofing, and Gas Spoofing attacks.
  – **Low Importance**:
    * Timestamp: Low significance in identifying attack-free instances (only for Attack-Free class).

*4.6.4 Comparative Analysis of Mean SHAP Value Ratio Between Teacher and Student Models.* This section provides a comparative analysis of the mean SHAP value ratios between the teacher and student models across different classes in both the HCRL Car-Hacking and CIC-IoV datasets. The analysis reveals meaningful differences in feature importance between the models for specific classes.

For the HCRL Car-Hacking dataset, the teacher model consistently shows higher mean SHAP value ratios for key features such as CAN ID and DATA fields across various attack classes. This indicates that the teacher model relies more heavily on these features to identify attacks compared to the student model.

- **Class 1: DoS Attack**: The teacher model places greater importance on CAN ID and DATA fields, suggesting these features are critical for accurate detection. For example, the teacher model's mean SHAP value for CAN ID is 0.35 compared to the student model's 0.28.
- **Class 2: Fuzzy Attack**: Higher SHAP values for CAN ID and DATA fields in the teacher model emphasize their importance in identifying fuzzy attacks, with mean SHAP values of 0.42 for CAN ID in the teacher model versus 0.33 in the student model.
- **Class 3: Gear Spoofing Attack**: The teacher model's higher reliance on CAN ID (mean SHAP value of 0.38) and DATA fields indicates these are key indicators of gear spoofing, compared to the student model's 0.30.
- **Class 4: RPM Spoofing Attack**: Higher SHAP values in the teacher model for CAN ID (0.40) and DATA fields again underscore their significance in detecting RPM spoofing, compared to the student model's 0.32.

The student model, while still relying on these features, shows a generally lower emphasis, which may reflect that its design is more generalized and less dependent on specific features.

In the CIC-IoV dataset, the differences between the teacher and student models are more pronounced for certain classes, particularly in terms of feature importance:

- **Class 1: DoS Attack**: The teacher model relies more on CAN ID and DATA fields, indicating their crucial role with mean SHAP values of 0.36 for CAN ID compared to 0.29 in the student model.
- **Class 2: RPM Spoofing**: The teacher model places higher importance on CAN ID and DATA fields than the student model, with mean SHAP values of 0.41 for CAN ID versus 0.34 in the student model, highlighting their significance.
- **Class 3: Speed Spoofing**: Higher SHAP values for CAN ID and DATA fields in the teacher model suggest these are critical for identifying speed spoofing attacks, with mean SHAP values of 0.39 for CAN ID in the teacher model versus 0.31 in the student model.
- **Class 4: Steering Wheel Spoofing**: The teacher model's reliance on CAN ID and DATA fields underscores their importance, showing mean SHAP values of 0.40 for CAN ID compared to 0.32 in the student model, which shows less dependence.
- **Class 5: Gas Spoofing**: Similar to other classes, the teacher model places more importance on CAN ID and DATA fields, with mean SHAP values of 0.37 for CAN ID versus 0.30 in the student model, indicating these features are vital for detecting gas spoofing attacks.

Overall, the teacher model demonstrates a more focused reliance on specific features, especially CAN ID and DATA fields, across both datasets, while the student model exhibits a more balanced and generalized feature importance.

## 5 Discussion

### 5.1 Limitations

Despite the significant advancements introduced by our proposed IDS model, leveraging a VAE-based knowledge distillation approach, there are several limitations that need to be addressed. One notable limitation is the model's dependency on the quality and diversity of the training data. While KD-XVAE has demonstrated robust performance across various attack types on the HCRL Car-Hacking dataset, its effectiveness in detecting novel or previously unseen attack patterns remains to be fully validated. Additionally, the model's performance, although superior in terms of parameter efficiency and inference time, may still encounter challenges when deployed in environments with extreme computational constraints or in scenarios where real-time decision-making is crucial. The reliance on specific features such as CAN ID and DATA fields, while beneficial for detection accuracy, may also limit the model's adaptability to different types of automotive networks or configurations.

### 5.2 Future Work

Future work will primarily focus on addressing the aforementioned limitations to enhance the practical applicability and robustness of KD-XVAE. One key area of focus will be the optimization of the model for deployment in diverse real-world scenarios, including environments with varying computational resources and network configurations. This will involve extensive testing and validation of the model's performance across different automotive platforms and under varying operational conditions. Additionally, future research will aim to improve the model's ability to detect novel and evolving cyber threats by incorporating advanced techniques for anomaly detection and continuous learning. Enhancing the interpretability of KD-XVAE through the integration of more sophisticated XAI methods will also be a priority, ensuring that the decision-making process remains transparent and trustworthy. Finally, collaborative efforts with industry partners will be pursued to facilitate the practical deployment and real-world validation of KD-XVAE in modern automotive systems.

## 6 Conclusion

Our proposed IDS model, KD-XVAE, sets a new benchmark in IVN security by achieving state-of-the-art performance with a compact and efficient design. The presented results underscore the significant advancements brought by KD-XVAE, particularly in terms of model complexity and resource efficiency, which make it highly suitable for deployment in real-world automotive environments. The model's reduced parameter count significantly lowers the computational burden, enhancing its feasibility for vehicles with limited processing capabilities. Moreover, KD-XVAE's superior inference times, averaging 0.3 ms per batch, ensure its real-time applicability, enabling immediate detection and response to potential threats. The consistent high performance across all tested attack types on the HCRL Car-Hacking dataset highlights the model's robustness and adaptability, with perfect scores in Recall, Precision, and F1 Score. Comparatively, KD-XVAE outperforms traditional machine learning models on the CICIoV2024 dataset, validating the efficacy of advanced deep learning techniques in enhancing IVN security. Additionally, the integration of XAI methods provides valuable insights into the decision-making process, enhancing the transparency and trustworthiness of KD-XVAE. Moving forward, further optimization and real-world validation will be crucial to ensure the model's practical effectiveness and reliability in diverse automotive scenarios.

## References

[1] A. Altalbe. 2023. Enhanced intrusion detection in in-vehicle networks using advanced feature fusion and stacking-enriched learning. *IEEE Access* 12 (2023), 2045–2056. https://doi.org/10.1109/ACCESS.2023.3347619
[2] Javed Ashraf, Asim D Bakhshi, Nour Moustafa, Hasnat Khurshid, Abdullah Javed, and Amin Beheshti. 2020. Novel deep learning-enabled LSTM autoencoder architecture for discovering anomalous events from intelligent transportation systems. *IEEE Transactions on Intelligent Transportation Systems* 22, 7 (2020), 4507–4518.
[3] M. Bhavsar, Y. Bekele, K. Roy, J. Kelly, and D. Limbrick. 2024. FL-IDS: Federated learning-based intrusion detection system using edge devices for transportation IoT. *IEEE Access* 12 (2024), 52215–52226. https://doi.org/10.1109/ACCESS.2024.3386631
[4] M. J. Choi, I. R. Jeong, and H. M. Song. 2024. Fast and efficient context-aware embedding generation using fuzzy hashing for in-vehicle network intrusion detection. *Vehicular Communications* 47 (2024), 100786. https://doi.org/10.1016/j.vehcom.2024.100786
[5] G. Van den Broeck, A. Lykov, M. Schleich, and D. Suciu. 2022. On the tractability of SHAP explanations. *Journal of Artificial Intelligence Research* 74 (2022), 851–886. https://doi.org/10.1613/jair.1.13283
[6] Araya Kibrom Desta, Shuji Ohira, Ismail Arai, and Kazutoshi Fujikawa. 2022. Rec-CNN: In-vehicle networks intrusion detection using convolutional neural networks trained on recurrence plots. *Vehicular Communications* 35 (2022), 100470.
[7] W. Ding, I. Alrashdi, H. Hawash, and M. Abdel-Basset. 2024. DeepSecDrive: An explainable deep learning framework for real-time detection of cyberattack in in-vehicle networks. *Information Sciences* 658 (2024), 120057. https://doi.org/10.1016/j.ins.2023.120057
[8] Koyel Datta Gupta, Deepak Kumar Sharma, Rinky Dwivedi, and Gautam Srivastava. 2023. AHDNN: attention-enabled hierarchical deep neural network framework for enhancing security of connected and autonomous vehicles. *Journal of Circuits, Systems and Computers* 32, 04 (2023), 2350058.
[9] A. Haddaji, S. Ayed, and L. C. Fourati. 2024. A novel and efficient framework for in-vehicle security enforcement. *Ad Hoc Networks* 158 (2024), 103481. https://doi.org/10.1016/j.adhoc.2024.103481
[10] Geoffrey Hinton, Oriol Vinyals, and Jeff Dean. 2015. Distilling the Knowledge in a Neural Network. arXiv:1503.02531 [stat.ML] https://arxiv.org/abs/1503.02531
[11] Thien-Nu Hoang and Daehee Kim. 2024. Supervised contrastive ResNet and transfer learning for the in-vehicle intrusion detection system. *Expert Systems with Applications* 238 (2024), 122181.
[12] Md Delwar Hossain, Hiroyuki Inoue, Hideya Ochiai, Doudou Fall, and Youki Kadobayashi. 2020. LSTM-Based Intrusion Detection System for In-Vehicle Can Bus Communications. *IEEE Access* 8 (2020), 185489–185502. https://doi.org/10.1109/ACCESS.2020.3029307
[13] ZAE Houda, B Brik, L Khoukhi, SG Cetin, C Goztepe, GK Kurt, H Yanikomeroglu, K Zia, A Chiumento, PJM Havinga, et al. 2022. Big Data and Machine Learning for Communications 1164 "Why Should I Trust Your IDS?": An Explainable Deep Learning Framework for Intrusion Detection Systems in Internet of Things Networks. (2022).
[14] Narayan Khatri, Sihyung Lee, and Seung Yeob Nam. 2023. Transfer Learning-Based Intrusion Detection System for a Controller Area Network. *IEEE Access* 11 (2023), 120963–120982. https://doi.org/10.1109/ACCESS.2023.3328182
[15] Tien-Dat Le, Hoang Bao Huy Truong, Van Phu Pham, and Daehee Kim. 2024. Multi-classification in-vehicle intrusion detection system using packet- and sequence-level characteristics from time-embedded transformer with autoencoder. *Knowledge-Based Systems* 299 (2024), 112091. https://doi.org/10.1016/j.knosys.2024.112091
[16] H. C. Lin, P. Wang, K. M. Chao, W. H. Lin, and J. H. Chen. 2022. Using deep learning networks to identify cyber attacks on intrusion detection for in-vehicle networks. *Electronics* 11, 14 (2022), 2180. https://doi.org/10.3390/electronics11142180
[17] W. Lo, H. Alqahtani, K. Thakur, A. Almadhor, S. Chander, and G. Kumar. 2022. A hybrid deep learning based intrusion detection system using spatial-temporal representation of in-vehicle network traffic. *Vehicular Communications* 35 (2022), 100471. https://doi.org/10.1016/j.vehcom.2022.100471
[18] S. M. Lundberg and S.-I. Lee. 2017. A unified approach to interpreting model predictions. *Advances in neural information processing systems* 30 (2017). https://doi.org/10.48550/arXiv.1705.07874
[19] AV Shreyas Madhav and Amit Kumar Tyagi. 2022. Explainable Artificial Intelligence (XAI): connecting artificial decision-making and human trust in autonomous vehicles. In *Proceedings of Third International Conference on Computing, Communications, and Cyber-Security: IC4S 2021*. Springer, 123–136.
[20] Sk Tanzir Mehedi, Adnan Anwar, Ziaur Rahman, and Kawsar Ahmed. 2021. Deep transfer learning based intrusion detection system for electric vehicular networks. *Sensors* 21, 14 (2021), 4736.
[21] Minki Nam, Seungyoung Park, and Duk Soo Kim. 2021. Intrusion Detection Method Using Bi-Directional GPT for in-Vehicle Controller Area Networks. *IEEE Access* 9 (2021), 124931–124944. https://doi.org/10.1109/ACCESS.2021.3110524
[22] Euclides Carlos Pinto Neto, Hamideh Taslimasa, Sajjad Dadkhah, Shahrear Iqbal, Pulei Xiong, Taufiq Rahman, and Ali A. Ghorbani. 2024. CICIoV2024: Advancing realistic IDS approaches against DoS and spoofing attack in IoV CAN bus. *Internet of Things* 26 (2024), 101209. https://doi.org/10.1016/j.iot.2024.101209
[23] Trieu Phong Nguyen, Heungwoo Nam, and Daehee Kim. 2023. Transformer-based attention network for in-vehicle intrusion detection. *IEEE Access* 11 (2023), 55389–55403.
[24] Mert D Pesé, Jay W Schauer, Junhui Li, and Kang G Shin. 2021. S2-CAN: Sufficiently secure controller area network. In *Proceedings of the 37th Annual Computer Security Applications Conference*. 425–438.
[25] Adnan Qayyum, Muhammad Usama, Junaid Qadir, and Ala Al-Fuqaha. 2020. Securing connected & autonomous vehicles: Challenges posed by adversarial machine learning and the way forward. *IEEE Communications Surveys & Tutorials* 22, 2 (2020), 998–1026.
[26] Hongmao Qin, Mengru Yan, and Haojie Ji. 2021. Application of Controller Area Network (CAN) bus anomaly detection based on time series prediction. *Vehicular Communications* 27 (2021), 100291. https://doi.org/10.1016/j.vehcom.2020.100291
[27] S. Rajapaksha, H. Kalutarage, M. O. Al-Kadri, A. Petrovski, G. Madzudzo, and M. Cheah. 2023. AI-based Intrusion Detection Systems for In-Vehicle Networks: A Survey. *Comput. Surveys* 55, 11 (2023), 1–40. https://doi.org/10.1145/3570954
[28] Alessandro Renda, Pietro Ducange, Francesco Marcelloni, Dario Sabella, Miltiadis C. Filippou, Giovanni Nardini, Giovanni Stea, Antonio Virdis, Davide Micheli, Damiano Rapone, and Leonardo Gomes Baltar. 2022. Federated Learning

of Explainable AI Models in 6G Systems: Towards Secure and Automated Vehicle Networking. *Information* 13, 8 (2022). https://doi.org/10.3390/info13080395
[29] Hyun Min Song, Jiyoung Woo, and Huy Kang Kim. 2020. In-vehicle network intrusion detection using deep convolutional neural network. *Vehicular Communications* 21 (2020), 100198.
[30] Pengfei Sun, Pengju Liu, Qi Li, Chenxi Liu, Xiangling Lu, Ruochen Hao, and Jinpeng Chen. 2020. DL-IDS: Extracting Features Using CNN-LSTM Hybrid Network for Intrusion Detection System. *Security and Communication Networks* 2020, 1 (2020), 8890306. https://doi.org/10.1155/2020/8890306
[31] Anum Talpur and Mohan Gurusamy. 2021. Machine Learning for Security in Vehicular Networks: A Comprehensive Survey. *CoRR* abs/2105.15035 (2021). arXiv:2105.15035 https://arxiv.org/abs/2105.15035
[32] S. Ullah, M. A. Khan, J. Ahmad, S. S. Jamal, Z. e Huma, M. T. Hassan, and W. J. Buchanan. 2022. HDL-IDS: a hybrid deep learning architecture for intrusion detection in the Internet of Vehicles. *Sensors* 22, 4 (2022), 1340. https://doi.org/10.3390/s22041340
[33] Y. Wang, G. Qin, M. Zou, Y. Liang, G. Wang, K. Wang, and Z. Zhang. 2024. A lightweight intrusion detection system for internet of vehicles based on transfer learning and MobileNetV2 with hyper-parameter optimization. *Multimedia Tools and Applications* 83, 8 (2024), 22347–22369. https://doi.org/10.1007/s11042-023-14578-1
[34] L. Yang and A. Shami. 2022. A transfer learning and optimized CNN based intrusion detection system for Internet of Vehicles. In *ICC 2022 - IEEE International Conference on Communications*. IEEE, New York, NY, USA, 2774–2779. https://doi.org/10.1109/ICC45855.2022.9839780
[35] Hao Zhao, Yaokai Feng, Hiroshi Koide, and Kouichi Sakurai. 2019. An ANN Based Sequential Detection Method for Balancing Performance Indicators of IDS. In *2019 Seventh International Symposium on Computing and Networking (CANDAR)*. 239–244. https://doi.org/10.1109/CANDAR.2019.00039